\documentclass[%
 reprint,
superscriptaddress,
 amsmath,amssymb,
 aps,
floatfix,
]{revtex4-2}

\usepackage{graphicx}
\usepackage{dcolumn}
\usepackage{bm}
\usepackage{amsmath,amssymb}

\begin{document}

\preprint{APS/123-QED}

\title{Charge-ordered state satisfying the Anderson condition in LiRh$_2$O$_4$ arising from local dimer order}

\author{M. Shiomi}
\affiliation{Department of Applied Physics, Nagoya University, Furo-cho, Chikusa-ku, Nagoya, Aichi 464-8603, Japan}
\author{K. Kojima}
\affiliation{Department of Applied Physics, Nagoya University, Furo-cho, Chikusa-ku, Nagoya, Aichi 464-8603, Japan}
\author{N. Katayama}\email{katayama.naoyuki@b.mbox.nagoya-u.ac.jp}
\affiliation{Department of Applied Physics, Nagoya University, Furo-cho, Chikusa-ku, Nagoya, Aichi 464-8603, Japan}
\author{S. Maeda}
\affiliation{Department of Applied Physics, Nagoya University, Furo-cho, Chikusa-ku, Nagoya, Aichi 464-8603, Japan}
\author{J.A. Schneeloch}
\affiliation{Department of Physics, University of Virginia, Charlottesville, VA 22904, USA}
\author{S. Yamamoto}
\affiliation{Department of Physics, Chiba University, 1-33 Yayoi-cho, Inage-ku, Chiba 263-8522, Japan}
\author{K. Sugimoto}
\affiliation{Department of Physics, Keio University, 3-14-1 Hiyoshi, Kohoku-ku, Yokohama, Kanagawa 223-8522, Japan}
\author{Y. Ohta}
\affiliation{Department of Physics, Chiba University, 1-33 Yayoi-cho, Inage-ku, Chiba 263-8522, Japan}
\author{D. Louca}
\affiliation{Department of Physics, University of Virginia, Charlottesville, VA 22904, USA}
\author{Y. Okamoto}
\affiliation{Department of Applied Physics, Nagoya University, Furo-cho, Chikusa-ku, Nagoya, Aichi 464-8603, Japan}
\author{H. Sawa}					
\affiliation{Department of Applied Physics, Nagoya University, Furo-cho, Chikusa-ku, Nagoya, Aichi 464-8603, Japan}
\date{\today}

\begin{abstract}
We report on the charge-ordered structure of LiRh$_2$O$_4$ arising below the metal-insulator transition at 170 K. Structural studies using synchrotron X-rays have revealed that the charge-ordered states of Rh$^{3+}$ and Rh$^{4+}$ with dimerization are realized in the low-temperature phase below 170 K. Although the low-temperature ground state resembles that of CuIr$_2$S$_4$, a charge ordering pattern satisfying the Anderson condition is realized in LiRh$_2$O$_4$. Based on structural information such as the short-range order of dimers appearing above the transition temperature and the weakening of the correlation between rhodium one-dimensional chains appearing in the crystal structure, we argue that the Coulomb interaction plays an important role in determining the charge ordering patterns.
\end{abstract}

\maketitle


Structural symmetry breaking caused by the ordering of electronic multiple degrees of freedom coupled to the lattice, occurs ubiquitously in transition metal compounds \cite{Khomskii, orbital_physics, orbital_effect}. Examples include iron chalcogenides/pnictides superconductors \cite{1111, 122, 11} and materials forming exotic low-temperature orbital molecules \cite{CuIr2S4, MgTi2O4, LiRh2O4, AlV2O4, LiMoO2, Li033VS2, CsW2O6, hirai, Li2RuO3, LiVS2-1, LiVO2, AlV2O4-2}. Recently, it was found that symmetry breaking can occur locally, as a precursor, even above the transition temperature \cite{Li2RuO3_2, AlV2O4-2, Attfield, CuIr2S4_2, MgTi2O4_2, MgTi2O4_3, LiVS2-2, FeSe-1, FeSe-2}. Spinel compounds CuIr$_2$S$_4$ and MgTi$_2$O$_4$ undergo a metal to nonmagnetic insulator transition accompanied by a structural distortion with the formation of orbital molecules at low temperature \cite{CuIr2S4, MgTi2O4}. Although a regular pyrochlore lattice was expected above the transition, recent experimental results using the pair distribution function (PDF) analysis of the local structure have shown that tetragonal distortions appear in the short-range length scales due to the realization of fluctuating $d$-orbital degeneracy lifting (ODL) \cite{CuIr2S4_2, MgTi2O4_2}. ODL states do not only appear in special systems, but are universally found in transition metal compounds, linked to several important phenomena in condensed matter physics \cite{CuIr2S4_2, NaTiSi2O6,La2OFe2OS2,Mn,FeSe-1, FeSe-2}. For example, ODL states have been predicted in the superconductor FeSe due to local nematicity \cite{FeSe-1, FeSe-2}.

The spinel LiRh$_2$O$_4$ has a 4$d^{5.5}$ electronic configuration that is isoelectric to CuIr$_2$S$_4$, and shows a nonmagnetic insulating transition below 170 K \cite{LiRh2O4}. Earlier HAXPES measurements \cite{LiRh2O4-HAXPES} and PDF analysis \cite{LiRh2O4_local} suggested charge separation between the Rh$^{3+}$ and Rh$^{4+}$ ions, and the dimerization in the low-temperature phase. In LiRh$_2$O$_4$, a structural transition from cubic to tetragonal occurs at a higher temperature of 220 K. This phase transition is considered to be a band Jahn-Teller transition, where the three degenerate $t_{2g}$ band manifolds are split into two bands, one with stabilized $yz$ and $zx$-characters and the other with destabilized $xy$-character \cite{LiRh2O4}. Although the ODL state has not been identified in LiRh$_2$O$_4$, it was proposed that the low temperature dimers survive locally in the high-temperature phase through the band Jahn-Teller phase \cite{LiRh2O4_local}. Strong electron correlations, which is different from that in metallic spinels such as CuIr$_2$S$_4$, is believed to be key to the formation of such short-range dimer. Therefore, we can expect LiRh$_2$O$_4$ is different from conventional ODL systems. How the local dimers develop into a low-temperature ordered phase is an important question. Moreover, the crystal structure of the low-temperature phase has not been identified yet.

In this letter, we report the identification of a charge-ordered structure with dimerization in the low-temperature phase of LiRh$_2$O$_4$. From synchrotron powder X-ray diffraction (PXRD) experiments, we clarified that the correlations between one-dimensional chains of Rh extending in the $xy$-plane were weakened in the band Jahn-Teller phase. Hereinafter, we refer to the one-dimensional chains of Rh with $xy$-, $yz$-, and $zx$-characters as ``$xy$-chain", ``$yz$-chain", and ``$zx$-chain", respectively. Consequently, the local dimers on the $xy$-chain becomes weakly interacting between the $xy$-chains. Upon cooling, the fluctuating local dimers become ordered, and a charge-ordered state with dimers is realized in the same manner as the ground state of magnetite proposed by Anderson \cite{magnetite1, magnetite2}. We argue that such ordered states are formed once the dimer fluctuations are stabilized by Coulomb interactions.

LiRh$_2$O$_4$ powder samples were prepared according to the reference \cite{LiRh2O4}. PXRD were performed at the BL5S2 beamline of the Aichi Synchrotron equipped with a PILATUS 100 K at $E$ = 19 keV. RIETAN-FP was used for the Rietveld analysis \cite{RIETAN}, and VESTA was used for graphing \cite{VESTA}. High energy PXRD for PDF analysis were carried out at BL04B2 of SPring-8 equipped with a hybrid Ge and CdTe detector at $E$ = 61 keV \cite{BL04B2}. The reduced PDF $G$($r$) was obtained by the usual Fourier transform of the collected data, where the PDFgui package was used \cite{PDFGUI}.

\begin{figure}
\includegraphics[width=80mm]{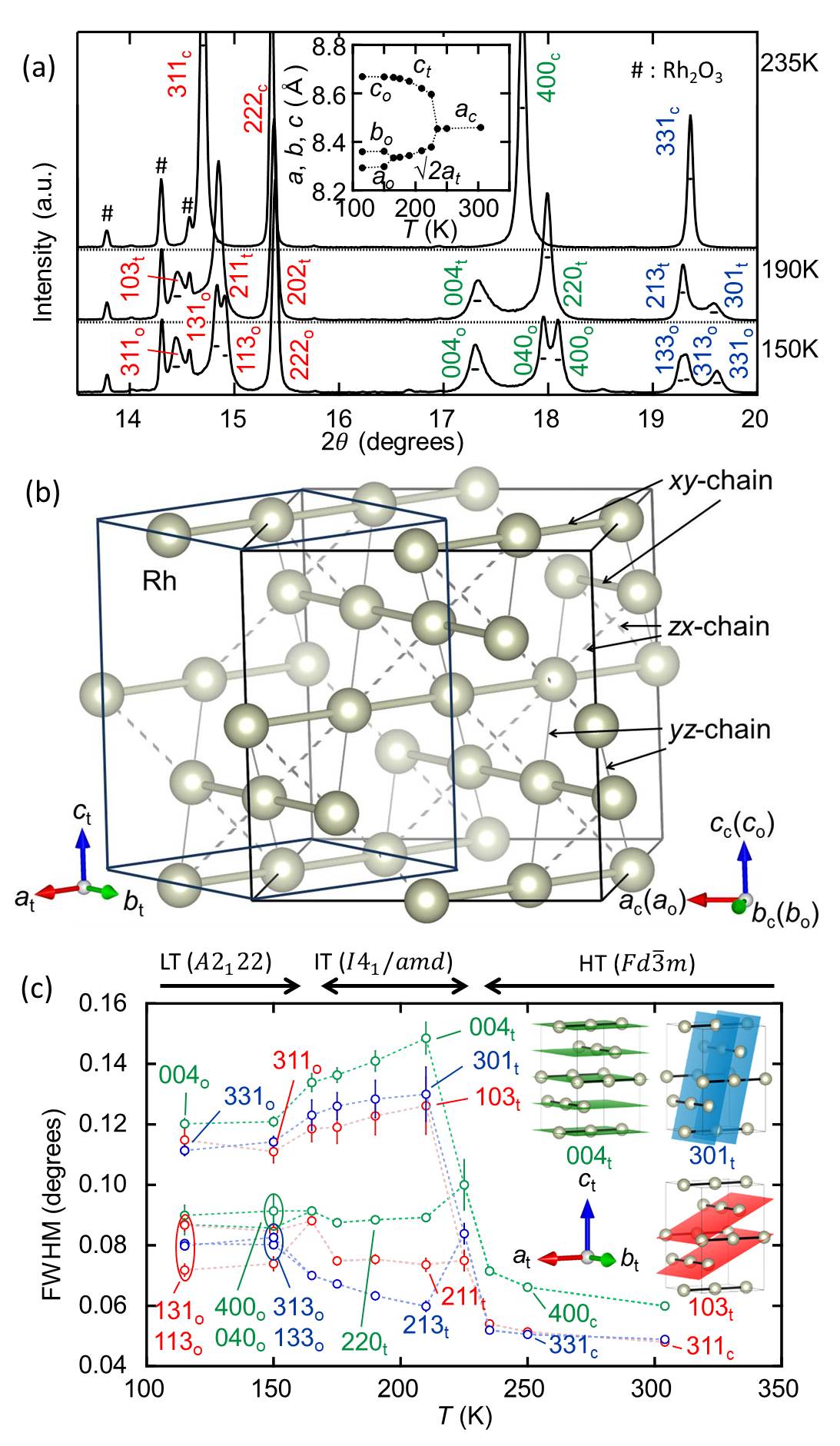}
\caption{\label{fig:Fig1} (a) PXRD patterns obtained at 235 K, 190 K and 150 K. The inset shows temperature dependent lattice parameters. (b) The unit cell and crystal axis in each temperature region with the $xy$, $yz$, and $zx$-chains consisting of Rh. (c) FWHM of some peaks colored in Figure~\ref{fig:Fig1}(a). The inset shows relationship between the $xy$-chains and the crystal planes corresponding to the colored indices.}
\end{figure}

Figure~\ref{fig:Fig1}(a) shows PXRD patterns of LiRh$_2$O$_4$ obtained in the high-temperature paramagnetic phase (HT-phase), band Jahn-Teller phase (IT-phase), and low-temperature nonmagnetic insulator phase (LT-phase), respectively. The diffraction data of the HT-phase was fitted assuming an undistorted cubic symmetry as summarized in Supplemental Information \cite{SI}, along with some Rh$_2$O$_3$ present as impurity (6.7\% in molar ratio). In all the following data, the analysis is performed assuming the existence of Rh$_2$O$_3$.  At 220 K, the band Jahn-Teller transition occurred, and the space group changes to $I4_1$/$amd$. The unit cell and the crystal axis are changed as shown in Figure~\ref{fig:Fig1}(b), where only the Rh ions are shown. Three Bragg peaks 311$_c$, 400$_c$, and 331$_c$, highlighted in red, green, and blue, respectively, are split into two peaks with different intensities. The lattice parameters obtained from the refinement are shown in the inset of Figure~\ref{fig:Fig1}(a).

In Figure~\ref{fig:Fig1}(c), the temperature dependence of the full-width half-maximum (FWHM) of the 311$_c$, 400$_c$, and 331$_c$ peaks of the HT-phase and their derivatives are shown. Of the two peaks that split and appear below 220 K, the weaker peak shows characteristic broadening. This broadening is slightly suppressed but maintained in the LT-phase. Such a peak broadening has already been reported \cite{LiRh2O4}, but the cause is not clear. Since broadening does not occur in all peaks, the broadening of the peaks does not originate from the short-range structural domains that occur in the band Jahn-Teller transition, but is thought to be due to the weakening of the long-range electron correlation between specific structural units inside the crystal. For example, the 00$l$ peak of a layered compound connected in the $c$-axis direction across the van der Waals gap can be broadened.

Since the atomic scattering factor of Rh is much larger than that of Li and O and contributes significantly to the diffraction intensity, we will focus on the contribution of Rh in the following sections and discuss the cause of the peak broadening. The pyrochlore lattice of LiRh$_2$O$_4$ consists of an entanglement of $xy$, $yz$, and $zx$-chains oriented by the $t_{2g}$ orbitals of Rh, as shown in Figure~\ref{fig:Fig1}(b). These three chains are crystallographically equivalent in the HT-phase, but in the IT-phase, the $xy$-chain becomes independent from the other. Figure~\ref{fig:Fig1}(c) inset shows the relationship between the $xy$-chains and the crystal planes of the 103$_t$, 004$_t$, and 301$_t$ indices in the IT-phase. In all of them, the $xy$-chain is always on the crystal plane, therefore, the interaction between neighboring $xy$-chains strongly affects the intensity and half-width of the Bragg peak of these indices. This kind of broadening was observed whenever the neighboring $xy$-chains were on the plane indicated by the index, suggesting that the electronic correlation between $xy$-chains in the IT-phase is weakened. Conversely, of the two peaks that split in the IT-phase, the sharp peaks are related to the $yz$- and $zx$-chains, not to the $xy$-chains, indicating that the long-range electronic correlation between these chains is preserved. The reason why only the correlation between the $xy$-chains is weakened is not clear, but it may be because the distance between the $xy$-chains increases with the band Jahn-Teller transition, or because the $yz$- and $zx$-bands, which reflect the electronic states of the $yz$- and $zx$-chains connecting the neighboring $xy$-chains, are occupied by electrons, leading to a weakening of the electrical interaction between the $xy$-chains via $yz$- and/or $zx$-chains.

Of the two split peaks present in the IT-phase, the sharper peak splits into two when the temperature is lowered to the LT-phase, as shown in Figure~\ref{fig:Fig1}(a). This indicates that the LT-phase is orthorhombic. Meanwhile, the 103$_t$, 004$_t$, and 301$_t$ peaks remain broad in the LT-phase, indicating that the short-range correlation between the $xy$-chains is maintained. Although the broadening of specific peaks is maintained, the following extinction laws can be found in the diffraction pattern of the LT-phase: $hkl$: $k$+$l$ $\neq$ 2$n$+1 and $h$00: $h$ $\neq$ 2$n$+1 when the crystal axes are taken in the same way for the HT- and LT-phases, as shown in Figure~\ref{fig:Fig1}(b). The orthorhombic space group satisfying these extinction laws is uniquely determined to be $A2_1$22. Hence, we identified the crystal symmetry of the LT-phase. 

\begin{figure}
\includegraphics[width=80mm]{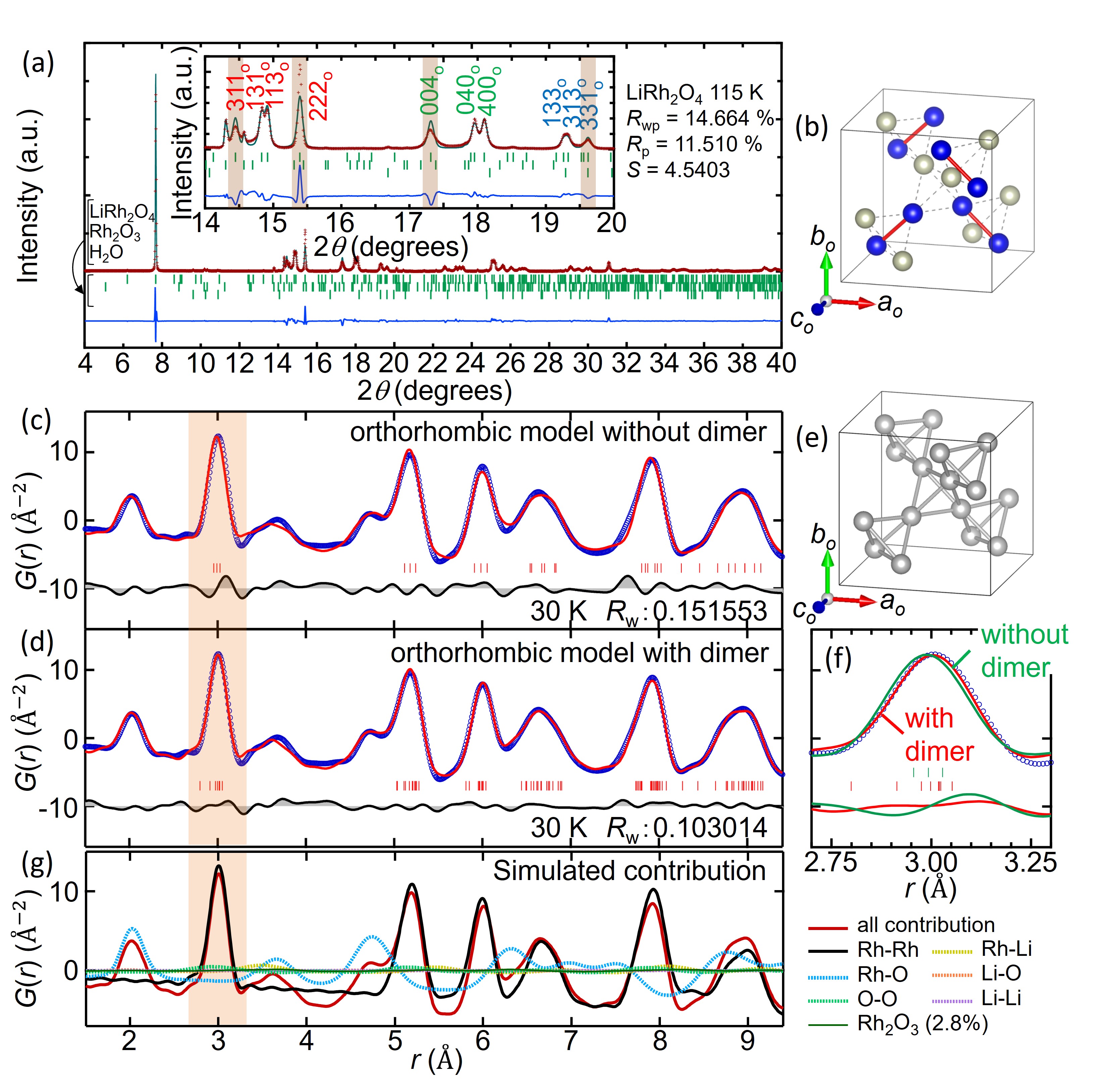}
\caption{\label{fig:Fig2} (a) Rietveld analysis for 115 K data using the space group $A2_1$22. The reliability factor $S$(= $R_{wp}$/$R_e$) is the ratio of the residual sum of squares, $R_{wp}$, to the smallest statistically expected $R_{wp}$, $R_e$. The peaks originating from the frost are refined together. Inset shows magnified view. The mixture of wide and sharp peaks makes it difficult to fit the peaks well, especially in the highlights. (b) Rh dimers revealed by Rietveld analysis. (c-d) Results of fitting the reduced $G$($r$) data with the models for (e) no displacement of Rh and (b) displacement, respectively. The tick marks indicate the Rh-Rh distance component. (f) Enlarged view around $r$(\AA) $\sim$ 3.0. (g) Simulated contribution calculated using the (b) model.}
\end{figure}


The results of the Rietveld analysis assuming space group $A2_1$22 are shown in Figure~\ref{fig:Fig2}(a). The ordered structure of Rh is shown in Figure~\ref{fig:Fig2}(b), confirming the formation of the dimer. The details of the refined structure are summarized in supplemental information \cite{SI}. The relatively large reliability factor $S$ is due to the persistence of peak broadening even at low temperatures. When sharp and broad peaks are mixed, as shown in the inset of Figure~\ref{fig:Fig2}(a), the width and intensity are not well refined for all peaks, resulting in large residuals and large $S$ values.

\begin{figure}
\includegraphics[width=80mm]{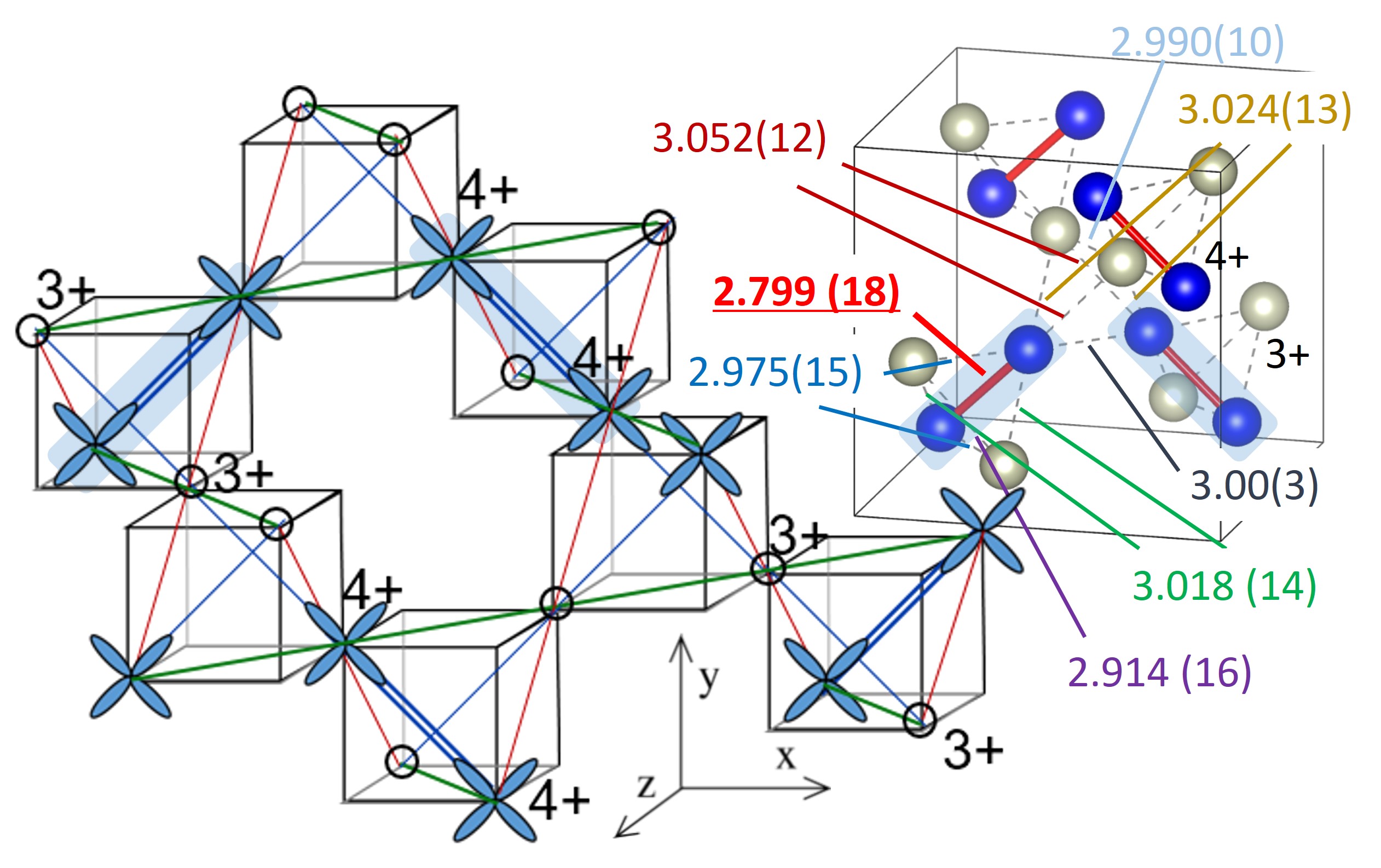}
\caption{\label{fig:Fig3} Charge-ordered pattern with dimerization in LT-phase, drawn in the similar fashion presented in reference \cite{orbitally}, and the Rh-Rh distances obtained from the analysis in Figure~\ref{fig:Fig2}(d). }
\end{figure}

The $G$($r$) corresponding to the local structure was obtained for the PDF analysis of the diffraction data. The data (blue circles) at 30 K are compared to a model $G$($r$) calculated using the $A2_1$22 symmetry in Figures~\ref{fig:Fig2}(c) and (d). First, in Figure~\ref{fig:Fig2}(c), the fitting was performed using the model shown in Figure~\ref{fig:Fig2}(e) without taking into account the atomic displacement of Rh. The simulation peak around $r$(\AA) $\sim$ 3.0 attributed to the nearest Rh-Rh distance is shifted to lower $r$ side relative to the experimental peak, indicating that there is a component with a longer Rh-Rh distance that is not considered in the current model. Assuming the presence of a short Rh-Rh bond associated with dimer formation, a large number of long Rh-Rh components appear as compensation. Assuming the dimer with an interdimer distance of 2.799 \AA, a large number of ticks appear around $r$(\AA) $\sim$ 3.0 as compensation, resulting in good fitting as shown in Figure~\ref{fig:Fig2}(f). The fitting around $r$(\AA) $\sim$ 5.2 and 7.8 are also improved. Note that these peaks are almost entirely due to the Rh-Rh distance, as shown in Figure~\ref{fig:Fig2}(g). Our results thus show that atomic displacements of Rh toward dimerization occur in the low-temperature phase.

The local structure we obtained is the only model that can explain charge separation and dimer formation in the $A2_1$22 space group, where Rh splits into two crystallographically equivalent 8$c$ sites. If dimer formation occurs between the different sites, all atoms should be involved in dimer formation and no charge separation should occur, which is inconsistent with the HAXPES results \cite{LiRh2O4-HAXPES}. Considering that the Rh$^{3+}$ ion is in the Jahn-Teller inactive $d^6$ electronic state and that the Rh$^{4+}$ ion is involved in dimer formation, the white and blue spheres can be safely assigned to Rh$^{3+}$ and Rh$^{4+}$, respectively in the unit cell models. Note that the charge separation was also supported from the Bond Valence Sum, as shown in the Supplemental Information \cite{SI}. Oxygen coordinates determined from PDF analysis using neutron diffraction data are used, and the details are summarized there as well.

The proposed structure is schematically shown in Figure~\ref{fig:Fig3} with the Rh-Rh distances. This is similar to the model used for the charge-ordered structure of the metallic spinel CuIr$_2$S$_4$ based on an orbitally induced Peierls transition \cite{orbitally}. Given the strong electron correlation of LiRh$_2$O$_4$ compared to CuIr$_2$S$_4$, a different mechanism is expected for charge ordering. The obtained charge-ordered structure with dimer is similar to that of CuIr$_2$S$_4$, but the charge-ordered arrangement occurring in the $xy$-chains is in a different phase between the $xy$-chains. This is very important because in the resulting charge-ordered arrangement of LiRh$_2$O$_4$, the sum of the number of charges inside any Rh$_4$ tetrahedron forming a pyrochlore lattice is equal, which is the pattern proposed by Anderson as the ground state of magnetite \cite{magnetite1, magnetite2}. It was proposed that the charge separation of Fe$^{2+}$ and Fe$^{3+}$ occurs as a consequence of the Coulomb interaction between cations, so that the number of charges inside any Fe$_4$ tetrahedron is equal. Although the ground state of magnetite has already been found to be different from that proposed by Anderson \cite{magnetite3, magnetite4, magnetite5, magnetite6, Radaelli}, the fact that the charge ordering pattern of LiRh$_2$O$_4$ satisfies the Anderson condition seems to suggest that Coulomb interaction plays a dominant role in the determination of the ground state.

\begin{figure}
\includegraphics[width=80mm]{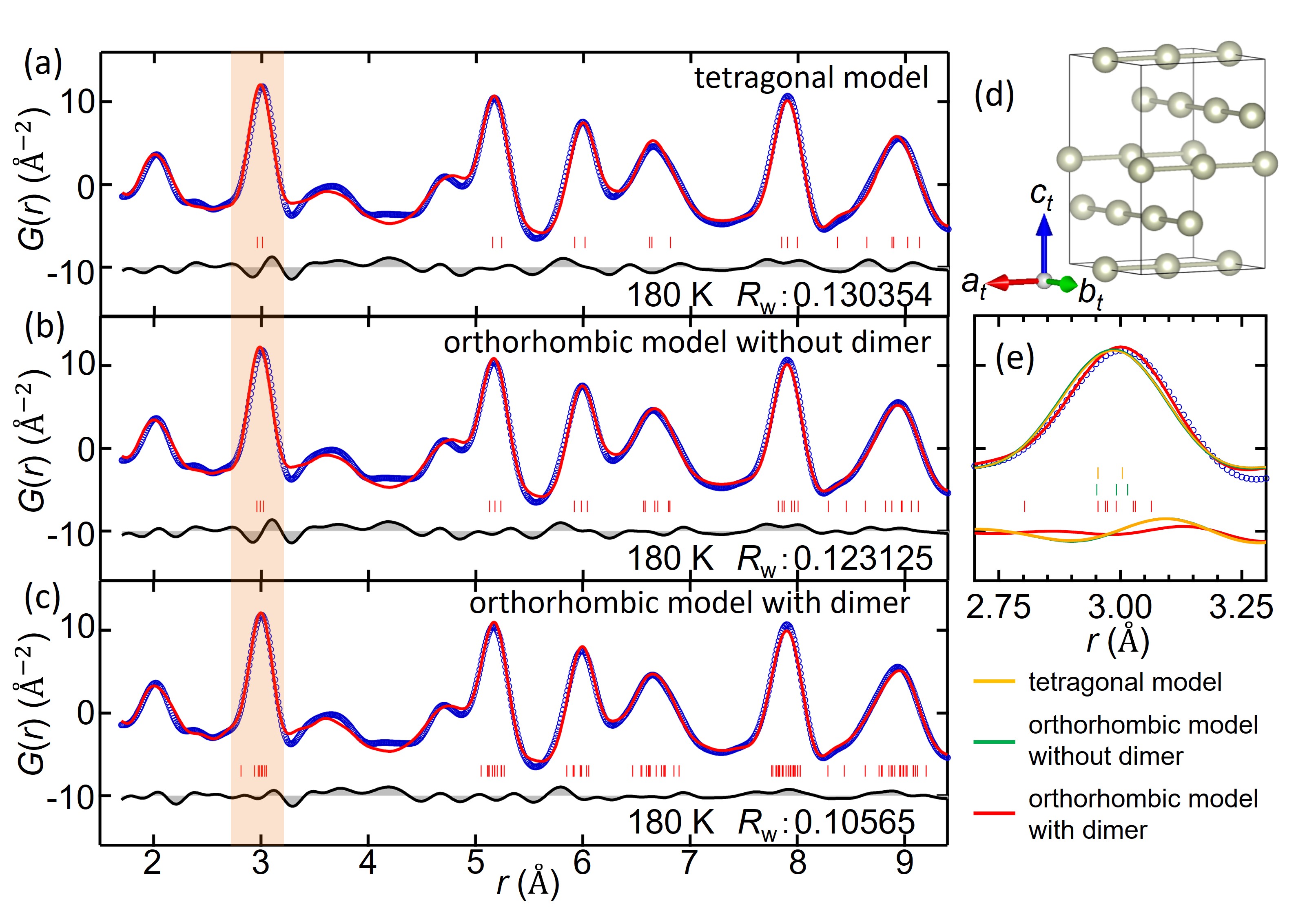}
\caption{\label{fig:Fig4} (a)-(c) Results of fitting reduced $G$($r$) data with  tetragonal model (d),  orthorhombic model without atomic displacement of Rh, and orthorhombic model with atomic displacement of Rh for dimerization, respectively. (e) Enlarged view around $r$ $\sim$ 3.0 for (a)-(c).}
\end{figure}

For our system, the local dimers forming the IT-phase reported previously may provide important information \cite{LiRh2O4_local}. Their presence was confirmed by fitting the data at 180 K in the IT-phase. Figure~\ref{fig:Fig4}(a)-(c) show the results of fitting to the data at 180 K with three different models. When fitting the data using the average structure of the IT-phase as shown in Figure~\ref{fig:Fig4}(d), a characteristic residual appears around $r$(\AA) $\sim$ 3.0  as in Figure~\ref{fig:Fig4}(e), which is similar to the one shown in Figure~\ref{fig:Fig2}(f). Although this residual survives when we assume an orthorhombic model without atomic displacements, it disappears when we assume an orthorhombic model with dimers.

If the dimers persists in the IT-phase but fluctuating with short-range ordering, this dimer fluctuation should be accompanied by charge fluctuations as well. Note that the results of previous HAXPES measurements supports this idea, since the Rh 3$d_{5/2}$ HAXPES data consists of Rh$^{3+}$ and Rh$^{4+}$ components even in the IT-phase \cite{LiRh2O4-HAXPES}. That is, in the IT-phase, the dimer fluctuations are accompanied by charge fluctuations in the $xy$-chain, and the strong electronic interaction between the $xy$-chains is absent. The remaining important factor causing charge ordering at low temperatures is the far-field force, the Coulomb interaction. As a result, the type of charge-ordered state proposed by Anderson as the ground state of magnetite is expected to be realized in the LT-phase.

Thus, in LiRh$_2$O$_4$, dimer fluctuations appearing in the IT-phase develop into the ordered state at LT-phase. Since the orbital degrees of freedom have already been lost in the IT-phase, which is a unique feature of LiRh$_2$O$_4$, it is clear that this dimer fluctuation is a different state from the ODL. Furthermore, it has been argued that the ODL is not a direct reflection of the low-temperature ordered state and needs to be studied independently \cite{CuIr2S4_2}, but the dimer fluctuations in LiRh$_2$O$_4$ seem to appear as a short-range order of the low-temperature ordered state. Such dimer fluctuations can be realized in many systems as well as in the ODL and significantly affect the ordering of the low-temperature phase.

Finally, we discuss the origin of the dimer fluctuations. For example, the ODL appears due to local nematicity in the iron-based superconductor FeSe. It has been argued that the existence of ODL explains the discrepancy between the tetragonal-orthorhombic transition temperature of 90 K and the huge energy splitting between 3$d_{yz}$ and 3$d_{xz}$ orbitals corresponding to 580 K \cite{FeSe-1, FeSe-2}. It has also been pointed out that a similar argument may be universally applied to many systems that undergo structural phase transitions at low temperatures \cite{CuIr2S4_2}. It remains to be seen whether the dimer fluctuations make an important contribution to the physical properties as in the ODL, but we note that the dimer fluctuations are reminiscent of the Cooper pair in superconductivity. In fact, superconductivity often appears in Rh spinel compounds with the same $d^{5.5}$ electric state as LiRh$_2$O$_4$ \cite{SC_rhodates, SC_rhodates2}. This may lead to a new research field of exotic superconductivity related to dimer fluctuations.



\begin{acknowledgments}
The work leading to these results has received funding from the Grant in Aid for Scientific Research (Nos.~JP17K17793, JP20H02604, 	JP21K18599). The work at the University of Virginia was partially supported by the Department of Energy, Grant No. DE-FG02-01ER45927. This work was carried out under the Visiting Researcher’s Program of the Institute for Solid State Physics, the University of Tokyo, and the Collaborative Research Projects of Laboratory for Materials and Structures, Institute of Innovative Research, Tokyo Institute of Technology. PXRD experiments were conducted at the BL5S2 of Aichi Synchrotron Radiation Center, Aichi Science and Technology Foundation, Aichi, Japan (Proposals No. 201801001, No. 201901017, No. 201901018, No. 201903064 and No. 2020L6002), and at the BL02B2 and BL04B2 of SPring-8, Hyogo, Japan (Proposals No. 2020A1059, No. 2019B1072, No. 2019B1073 and No. 2019A1218). Neutron diffraction experiments were conducted at the NOMAD beamline (BL-1B) of the Spallation Neutron Source at Oak Ridge National Laboratory (Proposal No. 27811). Note that the experimental details are summarized in Supplemental Information \cite{SI}
\end{acknowledgments}

\appendix

\nocite{*}

\bibliography{references}

\end{document}